\documentclass[prx,twocolumn,notitlepage,showpacs,amsmath,amstex,amssymb,citeautoscript,longbibliography,superscriptaddress,nofootinbib]{revtex4-1}
\pdfoutput=1
\usepackage{natbib}
\usepackage[english]{babel}
\usepackage{latexsym}
\usepackage[shortlabels]{enumitem}

\usepackage{graphicx}
\usepackage{amsmath}
\usepackage{mathtools}
\usepackage{amsfonts}
\usepackage{amssymb,bbding}
\usepackage{bm}
\usepackage{amssymb}
\usepackage{wasysym}
\usepackage{dsfont}
\usepackage{float}
\usepackage{braket}
\usepackage{subfiles}
\usepackage{epstopdf}
\usepackage[normalem]{ulem}
\usepackage{blkarray}
\usepackage{multirow}
\usepackage{hyperref}
\hypersetup{
    bookmarks=false,         % show bookmarks bar?
    unicode=false,          % non-Latin characters in AcrobatÃs bookmarks
    pdftoolbar=false,        % show AcrobatÃs toolbar?
    pdfmenubar=true,        % show AcrobatÃs menu?
    pdffitwindow=false,     % window fit to page when opened
    pdfstartview={FitH},    % fits the width of the page to the window
    pdftitle={Distinguishing localization from chaos: challenges in finite-size systems},    % title
    pdfauthor={D.A. Abanin, J.H. Bardarson, G. De Tomasi, S. Gopalakrishnan, V. Khemani, S.A. Parameswaran, F. Pollmann, A.C. Potter, M. Serbyn, and R. Vasseur},     % author
    pdfsubject={many-body localization},   % subject of the document
    pdfcreator={},   % creator of the document
    pdfproducer={}, % producer of the document
    pdfnewwindow=true,      % links in new window
    colorlinks=true,       % false: boxed links; true: colored links
    linkcolor=black,          % color of internal links (change box color with linkbordercolor)
    citecolor=blue,        % color of links to bibliography
    filecolor=magenta,      % color of file links
    urlcolor=blue           % color of external links
}

\newcommand{\be}{\begin{equation}}
\newcommand{\ee}{\end{equation}}
\newcommand{\bea}{\begin{eqnarray}}
\newcommand{\eea}{\end{eqnarray}}

\newcommand{\xxz}{XXZ$_{\rm nnn}$}

\begin{document}

\title{Distinguishing localization from chaos: challenges in finite-size systems}

\author{D.A. Abanin}
\affiliation{Department of Theoretical Physics, University of Geneva, 1211 Geneva, Switzerland}

\author{J.H. Bardarson}
\affiliation{Department of Physics, KTH Royal Institute of Technology, Stockholm, 106 91 Sweden}

 \author{G. De Tomasi}
\affiliation{Department of Physics, T42, Technische Universit\"at M\"unchen,
James-Franck-Stra{\ss}e 1, D-85748 Garching, Germany}

 \author{S.~Gopalakrishnan}
\affiliation{Department of Physics and Astronomy, CUNY College of Staten Island,
Staten Island, NY 10314, USA}
\affiliation{Physics Program and Initiative for the Theoretical Sciences,
The Graduate Center, CUNY, New York, NY 10016, USA}

 \author{V. Khemani}
\affiliation{Department of Physics, Stanford University, Stanford, California 94305, USA}
 
 \author{S.A.~Parameswaran}
\affiliation{Rudolf Peierls Centre for Theoretical Physics, Clarendon Laboratory, University of Oxford, Oxford OX1 3PU, UK}

 \author{F. Pollmann}
\affiliation{Department of Physics, T42, Technische Universit\"at M\"unchen,
James-Franck-Stra{\ss}e 1, D-85748 Garching, Germany}
\affiliation{Munich Center for Quantum Science and Technology (MCQST), Ludwig-Maximilians-Universit{\"a}t M{\"u}nchen, Fakult{\"a}t f{\"u}r Physik, Schellingstr. 4, D-80799 M{\"u}nchen, Germany}

\author{A.C.~Potter}
\affiliation{
 Department of Physics, University of Texas at Austin, Austin, TX 78712, USA}
\date{\today}

\author{M. Serbyn}
\affiliation{IST Austria, Am Campus 1, 3400 Klosterneuburg, Austria}
 
\author{R.~Vasseur}
\affiliation{Department of Physics, University of Massachusetts, Amherst, Massachusetts 01003, USA}
\date{\today}

\begin{abstract}
We re-examine attempts to study the many-body localization transition using measures that are physically natural on the ergodic/quantum chaotic regime of the phase diagram. Using simple scaling arguments and an analysis of various models for which rigorous results are available, we find that these measures can be particularly adversely affected by  the strong finite-size effects observed in nearly all numerical studies of many-body localization. This severely impacts their utility in probing the transition and the localized phase. In light of this analysis, we argue that a recent study [\v{S}untajs \emph{et al.}, \href{https://arxiv.org/abs/1905.06345}{arXiv:1905.06345}] of the behavior of the Thouless energy and level repulsion in disordered spin chains likely reaches misleading conclusions, in particular as to the absence of  MBL as a true phase of matter.
\end{abstract}

\maketitle

\vspace{1cm}

\section{Introduction}
 The investigation of non-equilibrium phenomena in quantum systems and their relevance to applications such as quantum computing is now an active research front in physics. Much theoretical and experimental work over the last decade has focused on the phenomenon of many-body localization (MBL) and its implications~\cite{Huse-rev,Altman-rev,AbaninRMP}. MBL, which generally requires strong quenched disorder, allows isolated quantum systems to evade thermalization. This frees MBL systems from certain limitations imposed by equilibrium statistical mechanics, opening a number of exciting opportunities. For example, MBL can protect quantum coherence and order in static and periodically driven systems, thereby extending the notion of phase structure to new, far-from-equilibrium regimes~\cite{Huse13L,Bauer13, Bahri:2015aa,ChandranKhemani}. Theoretically, MBL has been understood as a new phase of matter which exhibits robust emergent integrability~\cite{Serbyn13-1, Huse13, ScardicchioLIOM, imbrie2016many} that underpins its other unusual properties, such as absence of thermalization~\cite{PalHuse, OganesyanHuse, Znidaric08}, area-law entanglement of eigenstates~\cite{PalHuse, OganesyanHuse},  and logarithmic growth of entanglement in quantum quenches~\cite{Moore12,we,Znidaric08}.

A recent paper by \v{S}untajs, Bon\v{c}a, Prosen, and Vidmar~\cite{Vidmar2019} (henceforth SBPV) has claimed that MBL is not a phase of matter, but rather a finite-size regime that yields to ergodic behavior in the thermodynamic limit, i.e., when the system size $L\to\infty$. This conclusion was reached on the basis of a finite-size-scaling analysis of exact diagonalization (ED) studies of small ($L\leq 20$) one-dimensional (1D) spin models using diagnostics from quantum chaos --- the physical picture characterizing the ergodic, delocalized regime. Motivated by this striking claim, here we review and examine  existing theoretical and experimental work on MBL,
 focusing in particular on the finite-size scaling of various diagnostics used to probe the MBL transition.  
 
   \begin{figure}[t]
	\includegraphics[width=1.0\columnwidth]{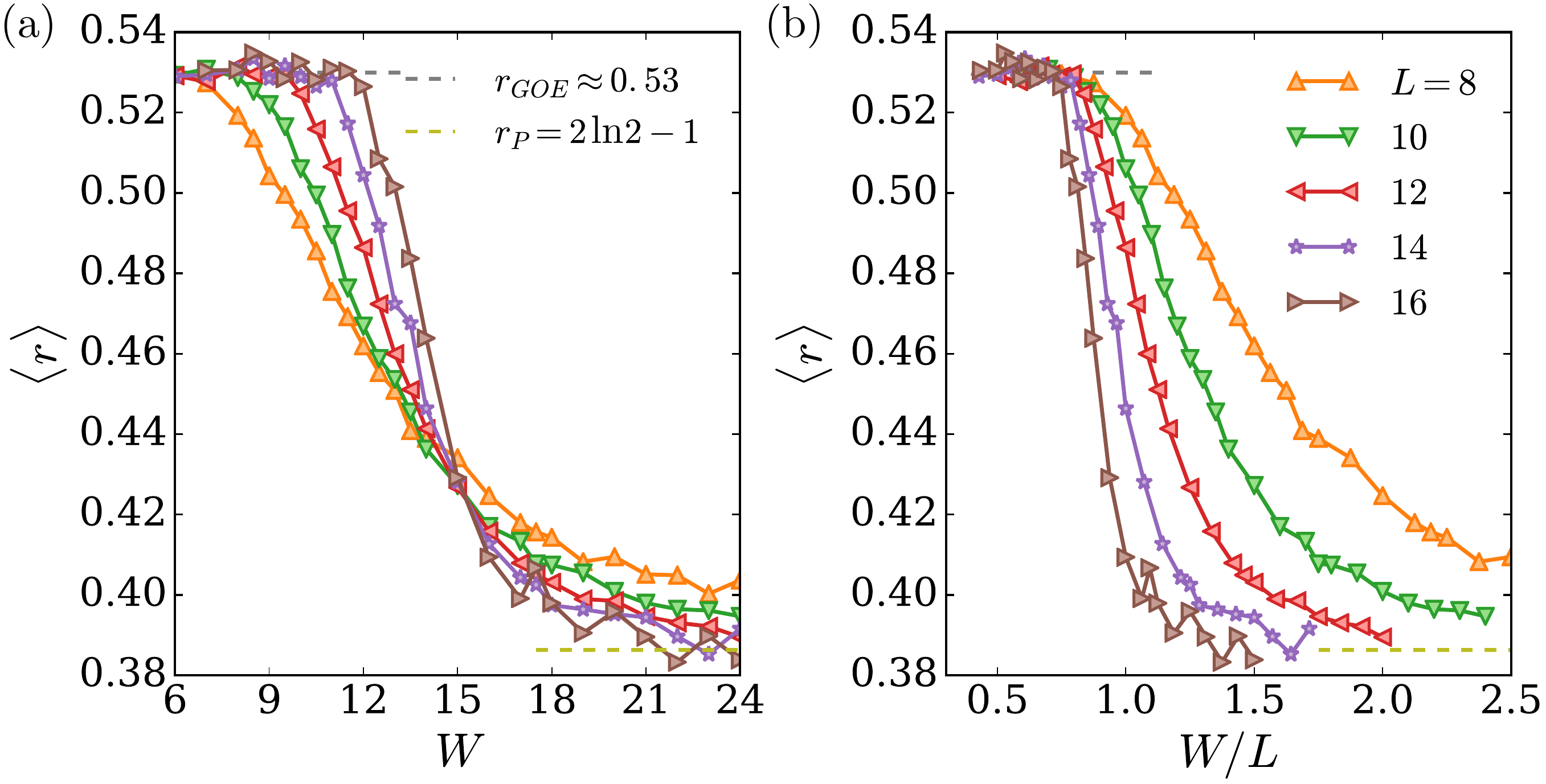}
	\caption{Level statistics {for the Anderson model on a random regular graph (RRG) with branching ratio $K=2$ and local connectivity $K+1=3$ (see Eq.~\ref{Eq:H_RRG}), plotted against disorder strength $W$ for different `system sizes' $N=2^L$.} (a) The dependence on the disorder- and spectrally-averaged ratio $\langle r\rangle$ of the minimum to the maximum of consecutive energy level spacings  of 32 states {in the middle of the spectrum} changes from the ergodic (`GOE') value $r_{\text{GOE}}\approx 0.53$ to the localized (`Poisson') value $r_{\text{P}}= 2\ln{2}-1\approx 0.38$ as disorder is increased, capturing the Anderson localization transition known to occur in this model  at $W_{\text{AT}} = 18.1 \pm 0.1$. (b) Plotting the same data with disorder scaled by system size (as for the model studied in Ref.~\cite{Vidmar2019}) seems to suggest that the phase boundary drifts linearly with $L$, incorrectly ruling out a transition in the thermodynamic limit. {We average over $1000(500)$ disorder realizations for $L\le 14$ ($L=16$).}}
	\label{Fig:RRG-r}
\end{figure}

The rest of the paper is organized as follows:  in Sec.~\ref{sec:background} we summarize the evidence for MBL, and clarify the nature of perturbative and non-perturbative mechanisms for destabilizing the localized phase, and their manifestations in finite-size scaling. (Readers familiar with MBL or who wish to go straight to numerical results may skip this section.) In Sec.~\ref{sec:numerics}, we then turn to a discussion of existing numerical probes of MBL in microscopic models, focusing in particular on two diagnostics --- the Thouless energy and the energy level statistics --- whose behavior in chaotic systems is well understood. While these or closely related quantities have been  studied previously, the relevant scaling analyses have mostly been rooted in expectations on the {\it localized} side of the transition. Such approaches necessarily presuppose the existence of a localized phase whose properties can be well-captured by numerical studies as long as the system size is larger than the localization length. We complement those previous analyses with an approach from the ergodic side, espoused by SBPV  as the correct perspective from which to view the putative transition without assuming the existence of MBL at the outset.

 In doing so, {in Sec.~\ref{sec:numerics}} we are inevitably led to focus on the strong finite-size effects characteristic of numerical studies of MBL, which are particularly pronounced in the ergodic phase. We explain how these effects greatly exacerbate the usual  dangers of extrapolating small-scale numerics to the thermodynamic limit. In order to illustrate these dangers, we demonstrate  how  scaling collapses based on extrapolating ergodic behavior (diffusion) to the strong-disorder regime can lead to demonstrably incorrect conclusions in exactly solvable models that share some features with the MBL transition.
 
As we discuss in detail in Sec.~\ref{sec:relatedscaling}, such extrapolations can incorrectly suggest that localization is absent even in examples where it is firmly established. This is vividly captured by Fig.~\ref{Fig:RRG-r}, which shows the $r$-parameter --- the ratio of the minimum to the maximum of consecutive energy level spacings --- as a function of disorder strength $W$ for the problem of Anderson (single-particle) localization on the random regular graph (RRG)~\cite{TikhonovRRG, KravtsovRRG2018, Kravtsov15, AndreaDe14, BeraRRG18, BiroliRRG18, GiuRRG19, GiorgioRRG18}. The unscaled data{, Fig.~\ref{Fig:RRG-r}(a),} shows a transition from $r_{\text{GOE}}\approx 0.53$ indicating the level repulsion characteristic of ergodic systems, to $r_{\text{P}} = 2\ln{2} -1\approx 0.38$ which is the value for the Poisson-distributed levels (with no repulsion) expected in the localized phase. 

Simply scaling the disorder strength by the system size (taken as $L = \log_2 N$ with $N$ the number of sites) {\it\`a la} SBPV's analysis, Fig.~\ref{Fig:RRG-r}(b), suggests that there is a finite-size crossover at $W^*(L) \sim 0.9L$ beyond which the level statistics deviates from predictions of RMT. This leads to the conclusion that the transition shifts inexorably to stronger disorder with increasing system size, suggesting that a localized {\it phase} is absent and that the system remains ergodic for arbitrary $W$ in the thermodynamic limit. This is clearly erroneous given the fact that for $L\rightarrow \infty$ the RRG converges to a Bethe lattice, on which the self-consistent theory of  Abou-Chacra, Thouless and Anderson~\cite{Abou_Chacra_1973} becomes exact and reveals a metal-insulator transition at a finite disorder strength~$W_{AT}$.
 
{In Section~\ref{sec:numerics}} we present similar results for other observables and other solvable or well-studied models, to underline the subtleties of extrapolating to the thermodynamic limit. {Afterwards,} having summarized the current understanding of numerics near the MBL transition, and exemplified some of the unusual scaling behaviours reported, we  turn to the discussion of the SBPV~\cite{Vidmar2019} results in Sec.~\ref{sec:vidmar}. We briefly summarize their results, examine them in light of the preceding analysis, and discuss possible ways in which the two may be reconciled.  

Of course we must remain open to the possibility  that an extremely subtle effect, missed by all previous studies,  leads to quantum chaos  destabilizing MBL even at strong disorder. Note that such an effect must also stem from a loophole in proposed proofs of MBL~\cite{Imbrie14}. However, as shown by the examples studied in this paper, addressing this question requires a more  careful analysis of finite-size scaling than has hitherto been attempted, in order to avoid arriving at incorrect conclusions. Our analysis thus injects a note of caution into the use of  new diagnostics from the ergodic regime to address the MBL transition using numerical studies at small system sizes.
 
\section{Theoretical Background \label{sec:background}}

\subsection{Evidence for MBL}
Although the possibility of MBL was already envisioned in Anderson's pioneering work~\cite{Anderson58}, and tentatively explored in the 1980s~\cite{Anderson80}, its existence has only been firmly established over the past decade. A key first step was the analytical work of Refs.~\cite{Basko06,Mirlin05}, which employed perturbative, locator-type expansions~\cite{Anderson58} to demonstrate the stability of localization to sufficiently weak short-range interactions between particles. These conclusions have received support from extensive numerical studies~\cite{OganesyanHuse,PalHuse,Znidaric08, Moore12,Serbyn13-1,we, Kjall14, Alet14, Bera15, Serbyn15, Goold15, Serbyn-16, Yu16, KhemaniCP, Pekker17, Bardarson17}, the majority of which relied on the exact diagonalization of disordered spin chains of size up to $L=24$~\cite{Alet14}. At sufficiently strong disorder, key expected properties of MBL have been observed, including: (i) Poisson level statistics~\cite{PalHuse,Alet14} indicating the absence of level repulsion (which is  a diagnostic of chaos); (ii) area-law entanglement of excited eigenstates~\cite{Serbyn13-1,Bauer13, Kjall14}; (iii) logarithmic spreading of entanglement following a quantum quench~\cite{Znidaric08, Moore12,we}; and (iv)  localization of conserved charges and hence absence of transport~\cite{Kjall14, Znidaric08, Serbyn14, Chandran14}. Several early studies~\cite{PalHuse,Alet14,Serbyn15} noted that simulations of MBL suffer from especially pronounced finite-size effects, necessitating extreme caution when extrapolating numerical results to the thermodynamic limit $(L\to \infty)$.
 Although finite-size studies are believed to be reliable either deep in the MBL phase, where the localization length $\xi$ 
 is much smaller than the system size,\footnote{We note that there are multiple ways of defining localization length in the MBL phase. The localization length $\xi$ discussed here controls locality of the quasi-local unitary transformation that relates eigenstates and product states, see Sec.~II.C.2 of Ref.~\cite{AbaninRMP}. This localization length is expected to diverge at the MBL transition.\label{note1}}\label{snote1} or deep in the ergodic phase, where the eigenstate thermalization hypothesis (ETH)~\cite{DeutschETH, SrednickiETH, RigolNature} is well-satisfied, extrapolating from either of these regimes to the transition region is challenging. 

The most rigorous piece of theoretical evidence for MBL is the work of Imbrie~\cite{Imbrie14}, who proved the existence of the MBL phase in a particular spin model subject to sufficiently strong disorder. More precisely, this work establishes the existence of a complete set of local integrals of motion. This proof is {non-perturbative and} rigorous, up to a physical assumption of `limited level attraction', but is limited to one dimension in contrast to locator expansions which work in arbitrary dimension. (It is perhaps worth noting here that proving localization is a challenging enterprise: the first proof of 1D Anderson localization~\cite{Goldshtein1977,kunz1980} appeared nearly two decades after Anderson's original work!)

Following these theoretical developments, experiments with ultracold atoms~\cite{Schreiber15,Choi16,Lukin2018}, trapped ions~\cite{Smith16}, nuclear spins~\cite{Wei18}, and superconducting qubits~\cite{Roushan17,Xu18} have probed the dynamics of isolated systems with tuneable disorder and interactions. Experiments have been able to probe large systems well beyond ED (e.g., up to $L=200$ ultracold atoms in dimensions $d=1,2$), but only over a finite timescale naturally limited by external dephasing and atom loss. The observed dynamics was found to be consistent with the existence of an MBL phase, but the finite observation time does not allow one to make statements about the eventual fate of the system at extremely long times.

\subsection{Possible Destabilizing Mechanisms and `Avalanches'}
The combined evidence (locator expansions, rigorous results, numerical simulations, and experiments), gives strong support for the existence of a `fully' MBL phase in one-dimensional systems with short-range interactions. 
While locator expansions demonstrate the {\it perturbative} stability of MBL in all dimensions, they fail to account for non-perturbative rare-region effects which could potentially destabilize MBL in other contexts. De Roeck and Huveneers~\cite{Roeck17} proposed that rare locally thermal inclusions (regions with atypically weak disorder) can drive an `avalanche': by thermalizing nearby spins, such inclusions can grow and become more efficient, eventually thermalizing the whole system. Such rare thermal inclusions are not included in the locator expansions and are in this sense non-perturbative.   A central feature of Ref.~\onlinecite{Imbrie14}'s proof of MBL in 1D  is to treat  such rare regions on special footing and demonstrate rigorously that  they do not `proliferate' and drive thermalization for sufficiently strong disorder. In this sense, while Ref.~\onlinecite{Imbrie14}  uses  perturbative  arguments in {\it typical}  regions, it accounts for non-perturbative  effects in rare regions. (We  remark in passing that the controversy around MBL in $d\geq  2$ centres on the severity of these  rare region effects, although  this is not our  focus  here.)

To understand avalanche-driven delocalization, let us first consider a rare region consisting of $\ell$ consecutive sites whose on-site energies $\varepsilon_i$ are all within the hopping amplitude $J$ of each other. Assuming the $\varepsilon_i$  fluctuate on scale $W$, the probability of occurrence of such a region is
$$
p(\ell)\approx \left ( \frac{J}{W} \right)^{\ell}.
$$
The typical size of such a region in a system of size $L$ is set by taking $Lp(\ell_{\rm typ})\sim 1$,  yielding  
$$
\ell_{\rm typ}(W) \sim \frac {\ln{L}}{\ln(W/J)}.
$$
Let us suppose for a moment that there  was a critical size $\ell_c$ of rare region, such that regions with $\ell>\ell_c$  trigger an avalanche and restore ergodicity (e.g., in terms of level repulsion) in the whole system. It is natural to assume that $\ell_c\gg 1$; this is also supported by recent numerical studies~\cite{Eisert_avalanche}. Taking $\ell_{\rm typ}(W)\sim \ell_c$,  gives an estimate for the scaling of critical disorder strength  with system size as $W_c(L) \sim J L^{1/\ell_c}$. If such `typical' avalanches caused thermalization, then (given $\ell_c\gg 1$) we would expect  a strongly sub-linear finite-size drift in $W_c(L)$, and hence absence of the MBL phase in the thermodynamic limit. However, the  proof in Ref.~\onlinecite{Imbrie14} considers  precisely such `typical' avalanches, and in  effect demonstrates that they do  not actually drive thermalization unless the localization length is above a critical value. This effective localization length is, in turn, enhanced by the presence of rare regions. 

The interplay of these two effects was first considered in Ref.~\onlinecite{MullerRG}, and its implications further explored in Refs.~\onlinecite{MBLKT,MorningstarHuse}, where it was argued that avalanches would lead to Kosterlitz-Thouless (KT)-like scaling behaviour. However, such avalanche-induced delocalization leads to a finite-size scaling of the critical disorder  strength  $W_c-  W_c(L)  \sim  (\ln L)^{-2}$. To date this KT scaling has only been directly observed in phenomenological models, and indirectly in one numerical study~\cite{Herviou19}. This suggests that  rare regions are not effective in driving the transition on the small system sizes accessible to exact numerics. 
   
A separate route to {de}localization might be a loophole in Imbrie's proof~[\onlinecite{Imbrie14}]. The most obvious assumption that could break down is that of limited level attraction~(LLA).  While at present we do not have a clear picture of how the failure of the LLA assumption would manifest in scaling, we note that there is no clear physical mechanism that appears to violate this assumption: ergodic systems show level repulsion while localized systems show its absence (Poisson level statistics). Level {\it attraction} would appear to require some additional symmetry, but the simplest MBL systems (such as the model in Ref.~\onlinecite{Imbrie14}) do not enjoy any symmetries beyond energy conservation. Level attraction is physically implausible, and in any case, if the LLA assumption were false, the consequent strong level attraction would lead to starkly different spectral statistics from the conventional chaos that all existing numerical studies (including SBPV) observe. Hence, we do not explore this possibility further. 

\section{Finite-Size Scaling and its Challenges in MBL Systems\label{sec:numerics}}

\subsection{Spectral Diagnostics from Exact Numerics}
We first discuss diagnostics of the MBL phase which can be extracted from exact finite-size spectra of models proposed to show an MBL transition. Two related signatures of ergodicity familiar from studies of chaotic systems are the appearance of level repulsion and  the  diffusive transport of energy and other conserved quantities.  The first is  probed by the level statistics parameter $\langle r \rangle$ defined in the introduction, averaged over the spectrum. This quantity  was first studied  numerically in Refs.~\cite{PalHuse,Alet14}, which found that, as a function of $W$, $\langle r \rangle$ exhibited a smeared step between the value expected for the Gaussian Orthogonal Ensemble (GOE) $r_{\rm GOE}\approx 0.53$ (at weak disorder) and a Poisson value $r_{\rm P}\approx 0.39$ (at strong disorder). This step sharpened with increasing $L$, indicative of a phase transition. A crossing point was present, but drifted to strong disorder with increasing system size --- a first sign of the severe finite-size effects at the MBL transition. Similar drift was seen in other measures, including  the scaling of eigenstate entanglement and its fluctuations~\cite{Devakul15, Alet14, Kjall14}.

A second measure is the diffusive transport in the ergodic regime. This can be characterized by the Thouless energy $E_{\rm Th} \sim D/L^2$, which is the inverse of the time needed for a conserved charge to diffuse across a system of linear scale $L$ (also called the Thouless time).  {The} Thouless energy~$E_{\rm Th}$ can be extracted from the energy spectrum in various ways.  In the single-particle problem $E_{\rm Th}$ can be computed by placing the system on a ring subject to twisted boundary conditions and examining the curvature of the energy spectrum~\cite{Thouless72, Mirlin}.
For many-body systems such as disordered spin chains,  it can be extracted by studying the spectral functions  of local operators~\cite{Polkovnikov-rev}.
In this setting $E_{\rm Th}$ is taken as the energy scale at which the spectral function becomes approximately constant, as this corresponds to the inverse of the transport time through the system~\cite{Serbyn-16}. 

Another way to extract Thouless energy, used by SBPV, is to study the time dependence of the spectral form factor (SFF), which is defined as the Fourier transform of the two-point correlations in the energy spectrum,
\begin{equation}\label{Eq:Ktau}
K(\tau) = \sum_{i\ne j} e^{i(E_i - E_j)\tau},
\end{equation}
where the $E_i$ are the many-body eigenenergies. In chaotic many-body systems the  SFF exhibits a characteristic linear increase, $K(\tau) \propto \tau$ for $\tau \in [\tau_{\rm Th},\tau_{\rm H}]$~\cite{Haake2006} i.e. for times between the  Thouless time  $\tau_{\rm Th} \sim 1/E_{\rm Th}$ and the Heisenberg time {$\tau_{\rm H} \sim 1/\Delta $, here taken to be  the inverse of the mean level spacing, $\Delta = \langle E_{i+1} - E_i\rangle$, which grows as  $\sim 2^L$ for spin-$1/2$ systems without U(1) symmetry}. This ramp indicates the scales between which the energy levels  display the level repulsion characteristic of random-matrix theory (RMT) behaviour expected of chaotic systems --- here,  the Wigner-Dyson (WD) statistics of the Gaussian orthogonal ensemble (GOE).  

Note, however, that in Hamiltonian systems this requires a `smoothing' procedure to eliminate spectral edge effects and an unfolding procedure needed to eliminate the effect of smooth changes in the many-body density of states --- see, e.g., Ref.~\onlinecite{Vidmar2019}. These procedures potentially introduce  additional  subtleties beyond those intrinsically present in the problem. Such concerns are less relevant in Floquet systems since there is no spectral edge and the density of states is uniform. We also observe that $K(\tau)$ in Eq.~\ref{Eq:Ktau} is in general not a \textit{self-averaging} quantity~\cite{Prange97, ECKHARDT97, Braun_2015, Cotler2017, Pasturbook}. As such, the disorder average of the SFF could be dominated by rare events, making it difficult to {reliably} extrapolate numerical results. 

 It is expected --- and well-known in single-particle systems~\cite{Altshuler86} --- that the Thouless energy defined via matrix elements and that extracted from the  SFF carry the same physical information. In the context of many-body systems, this connection has been recently established in Ref.~\onlinecite{Friedman2019} in a solvable Floquet model. {It is also worth noting that in $d=1$, several numerical simulations of transport have observed apparently {\it subdiffusive} behavior at finite times in the delocalized, near-critical regime~\cite{BarLev14,Demler14,Luitz-fluc-16, znidaric16PRL, Luitz-subdiff, Bera17, Doggen18, Weiner19, Prelov17, Gazit16}. This has been interpreted as a Griffiths phenomenon~\cite{Griffiths69, Demler14, AltmanRG14, Gopa-15} caused by the appearance of `bottlenecks' --- exponentially rare  regions through which transport is exponentially slow --- that leads to an effective time-dependent diffusion constant $D(t) \sim t^{2/z -1}$, so that in time $t$ a conserved charge travels a distance  $x(t) \sim \sqrt{D(t)t} \sim t^{1/z}$. Within the Griffiths scenario, as disorder is increased, $z$ increases continually until its divergence signals the onset of localization. Whether subdiffusion truly exists in large systems, and whether the observed subdiffusion has to do with Griffiths effects~\cite{schulz2019}, are not rigorously established. However, it is generally seen in small-size numerical studies, even at weak disorder, where it is theoretically unexpected and appears to cross over to diffusion in the largest systems studied~\cite{znidaric16PRL, Bera17,Gdt19_strong_V,Frank_Poll16, Lezama19}. This evidently complicates the task of extrapolating from the diffusive regime.

 Both the spectral function and SFF methods have a crucial limitation: they only provide information on $E_{\text{Th}}$ in the ergodic regime. This is because the onset of localization is accompanied by the absence of level repulsion, as originally discussed by Thouless~\cite{Thouless72}. Thus, both these methods effectively measure the energy window in which level repulsion exists.  In the localized phase, the Thouless energy becomes much smaller than the level spacing~$\Delta$, scaling as $E_{\rm Th}\propto e^{-L/\kappa}$, where $\kappa$ is a localization length~(see footnote~\ref{note1} in Sec.~\ref{snote1}). The spectral function and SFF methods only allow one to estimate (with significant errors) when $E_{\rm Th}$ becomes of the order of level spacing $\Delta$, but do not give insight into the properties of the MBL phase. 
  
   \begin{figure}[t!]
	\includegraphics[width=1\columnwidth]{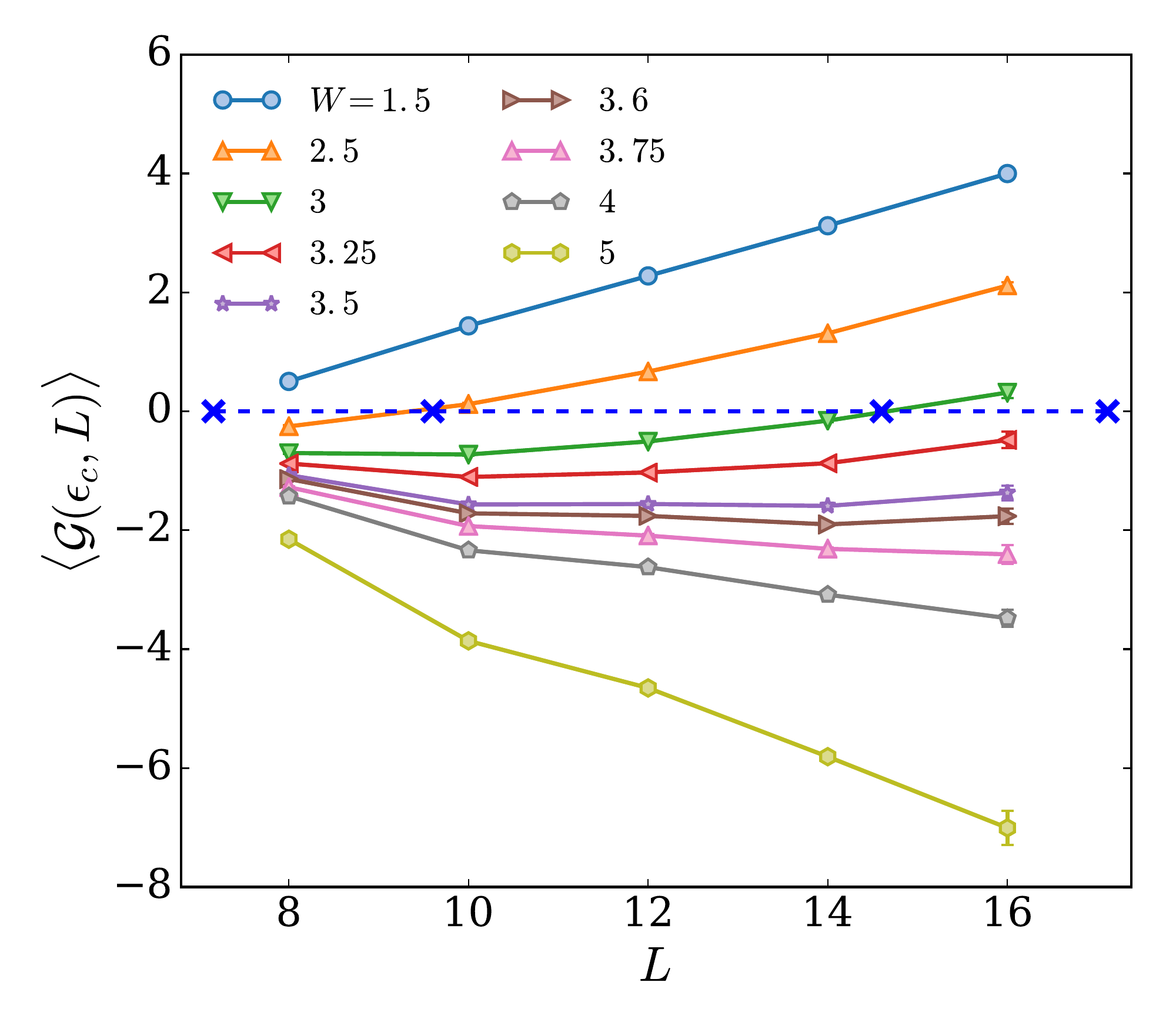}
	\caption{Finite-size effects in the matrix element-to-level-spacing measure $\mathcal{G}$. Note the strong non-monotonicity, indicating that estimates of the critical disorder strength would drift upward with increasing system size.   }
	\label{Fig:MatEl}
\end{figure}

An approach that gives similar information as the Thouless energy on the ergodic side but remains useful in the MBL phase is to study the behaviour of matrix elements of local operators, conveniently captured by ${\mathcal G}= \ln ({V_{\rm typ}}/{\Delta})$~\cite{Serbyn15}. This quantity has a simple physical interpretation: it gives the (log of the) probability that a typical local perturbation $V_{\rm typ}$  induces resonances by hybridizing many-body states differing  in energy by $\Delta \sim  e^{-L}$. $\mathcal{G}$ is expected to show  linear decay $\mathcal{G}(L)\propto -L$ above the MBL transition, consistent with the stability of MBL; such decay was indeed observed in numerics~\cite{Serbyn15}. 

However these studies found  very strong finite-size effects in the vicinity of the MBL transition:  $\mathcal{G}$ evolves non-monotically with $L$,  showing  an initial decay (as in the MBL phase) followed by an upturn and then the linear growth expected in the ergodic phase (Fig.~\ref{Fig:MatEl}).  When cut off by a small system size, this non-monotonic behaviour can lead to to an incorrectly small estimate of the critical disorder strength {$W_c$, given its drift with $L$}. We  note that such non-monotonicity is also characteristic of KT-like renormalization group (RG) flows for the MBL transition, where trajectories initially appear localized before eventually flowing to an  ergodic fixed point~\cite{MBLKT,MorningstarHuse}. It is {also observed in studies of localization on random regular graphs~\cite{Tikhonov19_2}}.

Such strong finite-size corrections are inevitable in numerical studies, even in the localized regime. Consequently, attempts to analyze the details of the MBL-ergodic transition (e.g., extracting critical exponents) based solely on ED studies have met with limited success. Phenomenological RG studies, including of solvable models, suggest that accessing the scaling regime requires very large system sizes. Evidence for this is bolstered by the fact that (unlike {phenomenological RG studies}) ED studies often yield exponents~\cite{Alet14,Serbyn15} inconsistent with general bounds~\cite{Harris, Harris2, CCFS, Chandran2015}.
 Furthermore, given the strong finite-size effects, even a seemingly innocuous change to a model (e.g, adding a longer-range hopping as in SBPV) can slow the rate at which non-universal contributions vanish as $L\rightarrow \infty$ and  thus worsen the scaling properties accessible via ED. Given these concerns it is natural to view numerical evidence for MBL with some caution, particularly near the transition. However, as we now show this issue is more fundamental: even in models with a well-{established} localized regime, an approach to scaling motivated from the ergodic side shows strong finite-size corrections that, interpreted naively, would indicate that a localized phase is absent. 

\subsection{Scaling in Related Problems: `Missing' Localized Phases\label{sec:relatedscaling}}
We now change gears and  consider three related problems, all of which share the feature that the existence of the localized phase is firmly established.
\begin{enumerate}[(i)]

\item The  Anderson  model on the RRG  (as reported in the introduction), given by
\begin{equation}\label{Eq:H_RRG}
H=  -\sum_{ x \sim y } \ket{x}\bra{y} + \sum_x \epsilon_x\ket{x}\bra{x}.
\end{equation}
Here, $\ket{x}$ denotes a site on the RRG, the sums range over $N=2^L$ sites, and `$\sim$' denotes sites that are adjacent on the RRG, which is taken to have a fixed branching number $K=2$, corresponding to a local connectivity of $K+1=3$. The $\{\epsilon_x\}$ are independent random variables distributed uniformly between  $[-W/2,W/2]$. Although there remains some debate over the existence of a non-ergodic but delocalized phase in this model~\cite{AndreaDe14, TikhonovRRG, Tikhonov19, BeraRRG18, GiuRRG19, BiroliRRG18,Sonner17, Tikhonov19_2, Metz17, GilRRG19}, it is known to have an Anderson localization transition at $W_{AT} \approx 18.1 \pm 0.1 $~\cite{GiorgioRRG18,KravtsovRRG2018, Tikhonov19, Abou_Chacra_1973, MIRLIN1991507}. This is consistent with the expectation, as noted in the introduction, that the self-consistent theory of localization~\cite{Abou_Chacra_1973} becomes exact for the RRG in the thermodynamic limit, where it predicts such a transition. 
\item The `Imbrie model', described by the Hamiltonian \be \label{eq:ImbrieHam} H=\sum_{i=1}^{L-1} J_i \sigma^z_i \sigma^z_{i+1}+ \sum_{i=1}^L\left( h_i \sigma_i^z + \gamma_i\sigma^x_i\right),\ee 
We fix $\gamma_i=1$, and choose the remaining couplings from uniform distributions, $J_i \in [0.8, 1.2]$, $h_i \in [-W,W]$, and study $L$-site chains with open boundary conditions . The existence of MBL in this model has been established rigorously (under the assumption of limited level attraction) in Ref.~\onlinecite{imbrie2016many}.
\item A family of phenomenological classical models with {an infinite randomness critical point} introduced to model the phenomenology of MBL transition via RG~\cite{AltmanRG14,Potter15X}. In particular, we consider a recently introduced solvable model~\cite{GoremykinaPRL} which is a deformation of a coarsening model that allows controlled access to a critical point. This one-dimensional model implements simple rules for how to merge adjacent `thermal' and `insulating' regions of randomly distributed lengths in a manner that can be studied via a real-space renormalization group approach. 
\end{enumerate}
  \begin{figure}[t]
	\includegraphics[width=1\columnwidth]{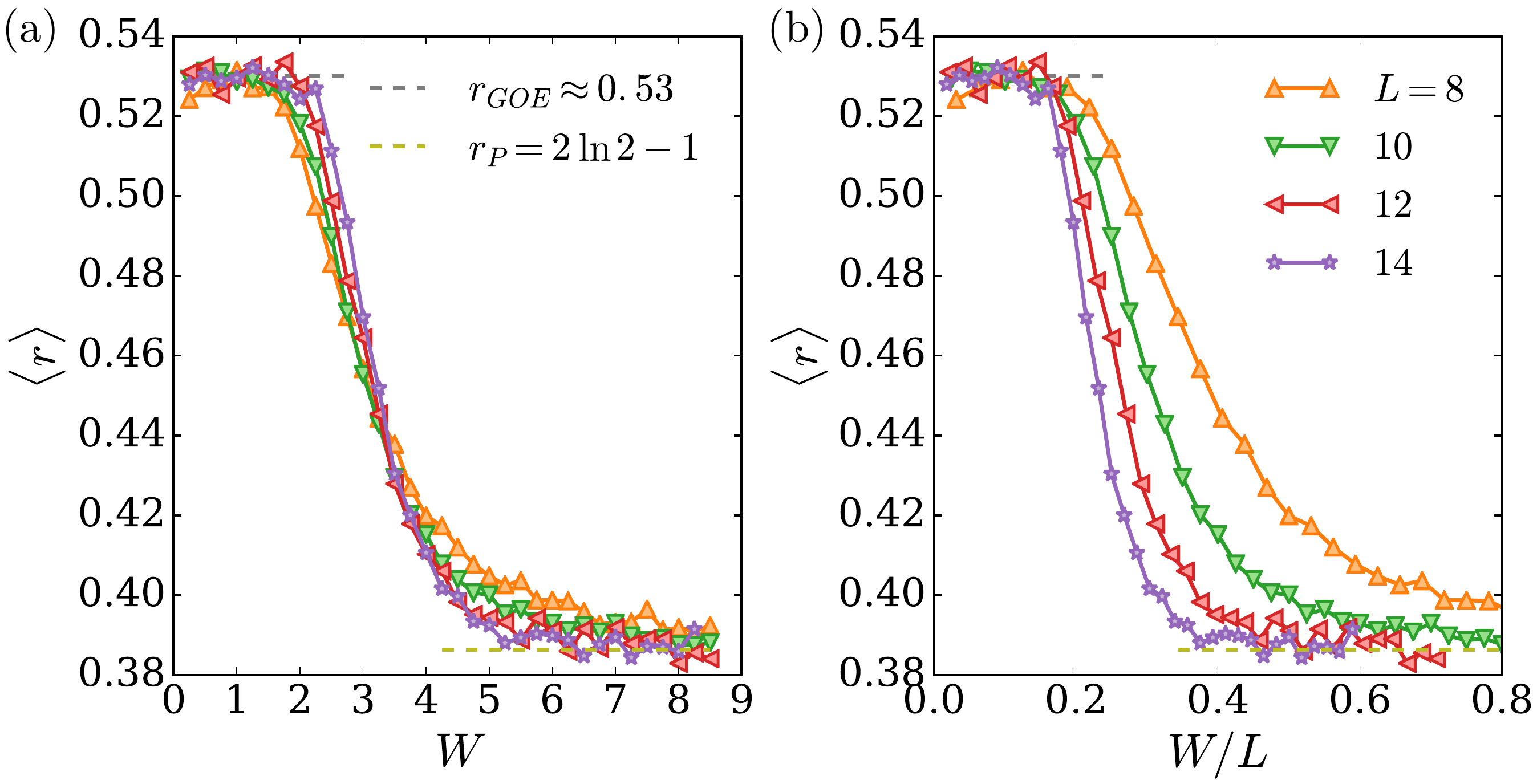}
	\caption{Level statistics for the `Imbrie model'. Similar finite-size effects to the RRG case reported in the Introduction (Fig.~\ref{Fig:RRG-r}) and also studied in SBPV for {the random-field XXX and \xxz\ spin chains} are observed: scaling the disorder strength by system size suggest a crossover moving to infinite disorder in the thermodynamic limit rather than a true phase transition. Localization has been proved in this model at strong disorder under the assumption of limited level attraction in Ref.~\cite{imbrie2016many}. We average over $1000$ disorder realizations and over $32$ eigenstates in the middle of the spectrum.}
	\label{Fig:Imbrie-r}
\end{figure}

Models (i) and (ii) are  microscopic and so it is possible to extract and characterize the energy spectrum as a function of disorder strength. The simplest quantity to compute is the $r$-ratio. The behaviour of this quantity for the RRG problem was illustrated in Fig.~\ref{Fig:RRG-r}  in the introduction (and was previously studied in Ref.~\onlinecite{TikhonovRRG}) and for the Imbrie model is shown in Fig~\ref{Fig:Imbrie-r}. In the RRG case, the role of the system size is played by the logarithm of the number of sites $N$: $L=\log_2 N$. In the spirit of  SBPV, we use the data to define the extent of the ergodic region by determining the disorder strength ${W}_{\rm erg}(L)$ where $\langle r \rangle$ first deviates from the value predicted by RMT. In each case,  the extent of the ergodic regime appears to grow with increasing system size, and a naive collapse with data yields $W_{\rm erg}\propto L$. This is despite the fact that the existence of localized phase is well-established~\cite{GiorgioRRG18, KravtsovRRG2018, Tikhonov19, Abou_Chacra_1973, MIRLIN1991507}. We therefore conclude that the apparent unbounded growth of the ergodic regime with system size for $L\leq 20$ is consistent with localization at $W>W_c$ with $W_c$ finite in the thermodynamic limit as necessary for a localized phase to exist. 

  \begin{figure}[t]
	\includegraphics[width=1\columnwidth]{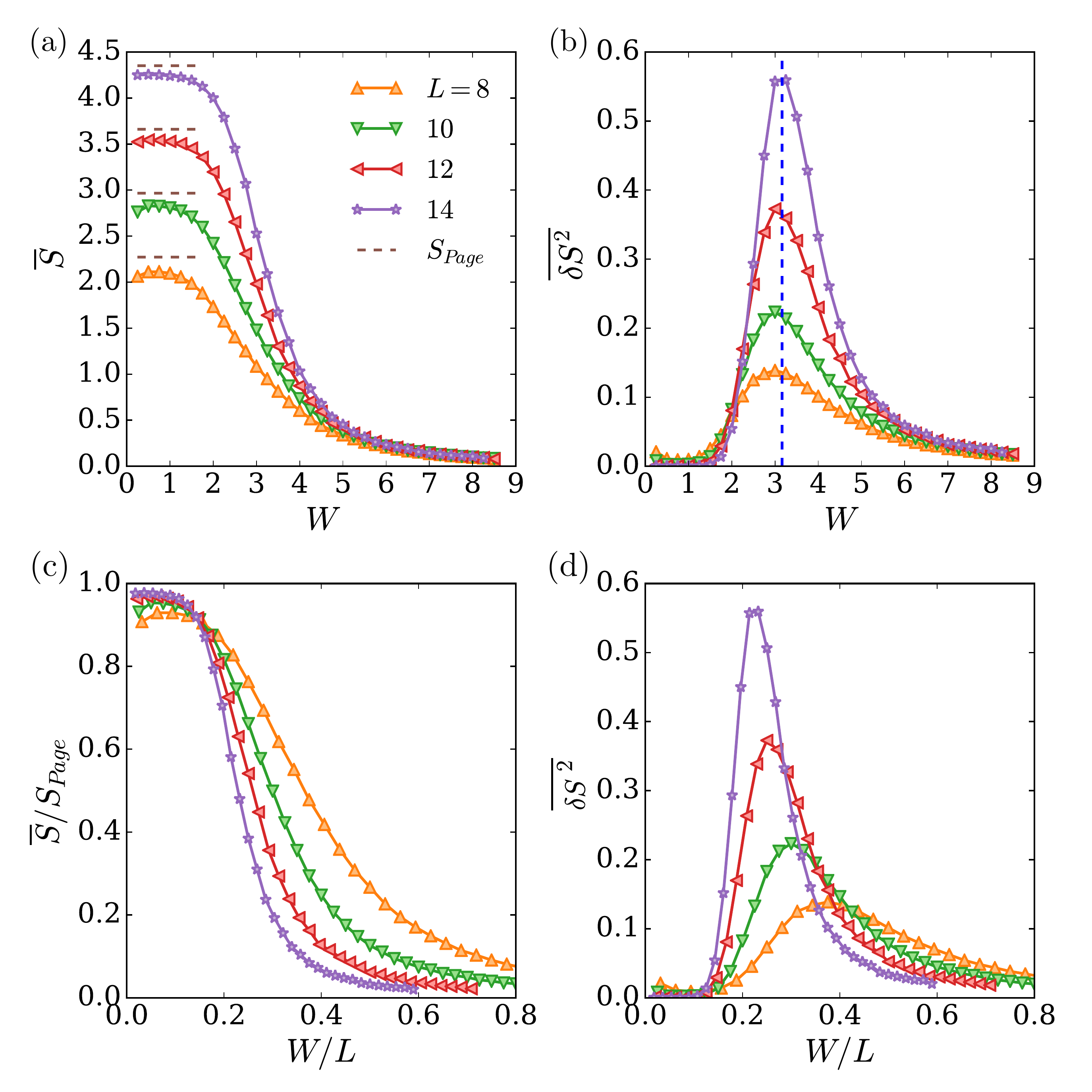}
	\caption{Entanglement entropy scaling in the `Imbrie model' [cf Eq.(\ref{eq:ImbrieHam})]. (a) The disorder-and eigenstate-averaged half-chain entanglement entropy $S = -\text{Tr} \rho_{1/2} \ln \rho_{1/2}$ where $\rho_{1/2} =\text{Tr}_{x>L/2} \ket{\Psi}\bra{\Psi}$ is the reduced density matrix of the left half of an $L$-site chain in eigenstate $\ket{\Psi}$, plotted against disorder strength $W$. This shows a transition from `volume law' scaling $S\propto L$ expected in the ergodic phase, to the area law behaviour $S\propto \text{const.}$ characteristic of MBL systems. The dashed lines indicate the averaged entanglement entropy for random pure states $S_{\text{Page}} = L/2 \ln{2}-1/2$. (b) Fluctuations $\overline{\delta S^2}$ in $S$, again plotted against $W$. The peak sharpens with increasing system size and can be taken as a proxy for locating the transition {(vertical dashed line)}. Panels (c)-(d) show  the same data as in (a)-(b) but now plotted against the scaled variable $W/L$ and with $y$-axis rescaled by Page value of entropy for (c). Note the finite-size drift in the data, which is consistent with the drift reported in the $r$-ratio for this model. We average over $1000$ disorder realizations and over $32$ eigenstates in the middle of the spectrum.}
	\label{Fig:Imbrie-S} 
\end{figure}

{For the Imbrie model, it is useful to also compare the behaviour of the $r$-ratio with the scaling of the bipartite, half-system eigenstate entanglement entropy. For a 1D system of length $L$ in eigenstate $\ket{\Psi}$, this is given by  $S = -\text{Tr} \rho_{1/2} \ln \rho_{1/2}$ where $\rho_{1/2} =\text{Tr}_{x>L/2} \ket{\Psi}\bra{\Psi}$ is the reduced density matrix of the left half of the system. We may average this quantity over eigenstates and over disorder realizations.
The eigenstate average will be dominated by `infinite temperature' states near the middle of the many-body spectrum, which for ergodic systems satisfy `volume law' scaling $S(L) \propto L$. On the localized side, MBL implies an area law for all but a measure-zero set of states in the spectrum, and so we expect $\bar{S}$ to scale as a constant with system size $S(L) \sim \mathcal{O}(L^0)$. Fig.~\ref{Fig:Imbrie-S}(a) clearly shows {a crossover} from volume- to area-law scaling with increasing disorder strength{, that sharpens for increasing $L$ consistent with a transition in the thermodynamic limit}. Fig.~\ref{Fig:Imbrie-S}(b) shows that the fluctuations $\overline{\delta S^2} = \overline{S^2} - \overline{S}^2$ of entanglement (the average is over both disorder and eigenstates) are maximal near the {putative} transition between the two scaling behaviors, underscoring the role played by entanglement in developing theories of the MBL transition. However, as shown in Fig.~\ref{Fig:Imbrie-S}(c)-(d), plotting the rescaled entanglement and entanglement fluctuations against the scaled disorder $W/L$ shows similar finite-size drift as the $r$-ratio, again indicating that this drift is {an apparently inevitable} feature of numerical studies of a transition. As we have seen from the study of the RRG, such drifts exist even in systems with a well-defined localized phase.
}

   \begin{figure}[t]
	\includegraphics[width=1\columnwidth]{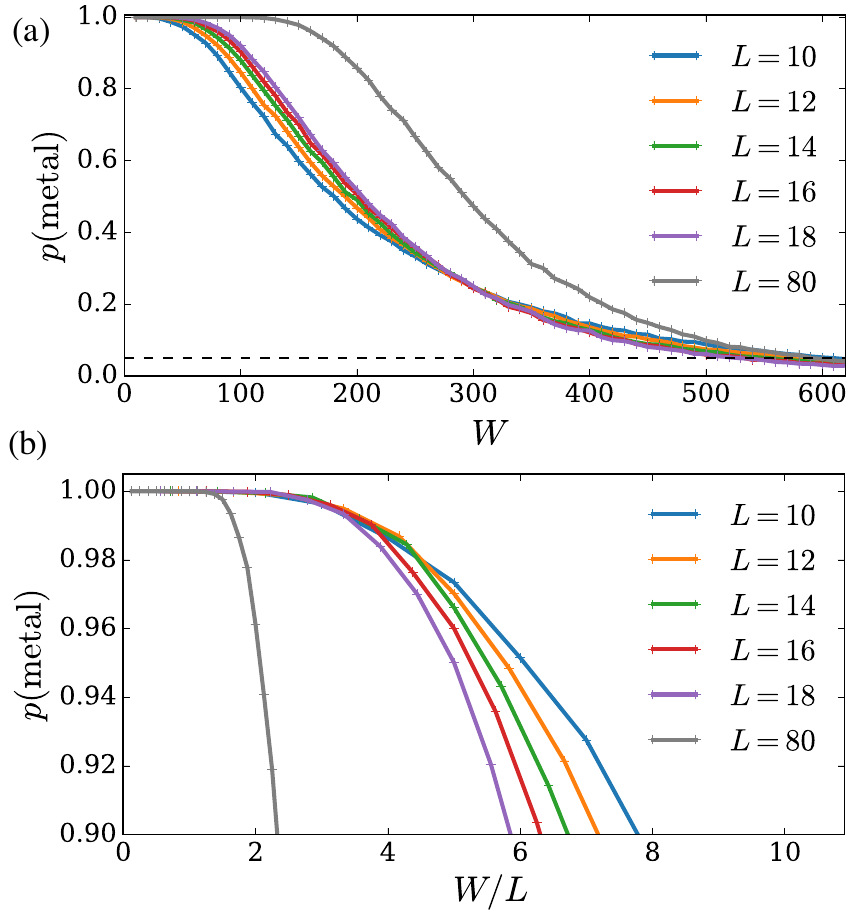}
	\caption{Finite-size effects in the solvable, phenomenological RG model of Ref.~\onlinecite{GoremykinaPRL} {for the probability for the system to thermalize}, {with the parameter choice $\alpha=1/20, \beta=20$}. At small system sizes these effects are similar to those seen in ED studies of microscopic MBL models. These sizes are far from the thermodynamic limit, as evident from in the figure (contrast data at $L=80$ against the rest: {the dashed horizontal line in the upper panel represents the exact probability to thermalize at criticality}). This model is known to exhibit two phases, but naive extrapolation along the lines of SBPV would predict only one phase in thermodynamic limit. {Averaging is performed over at least 20000 disorder realizations.}}
	\label{Fig:PhenoRG}
\end{figure}

{The RG models do not admit as direct a comparison as models (i) and (ii), since (being classical in nature) they do not have a notion of an eigenspectrum. Nevertheless, {the} existence of an analytical solution in the thermodynamic limit ensures the presence of a phase transition and allows one to obtain both its critical exponents as well as other physical characteristics, such as the probability that {the system thermalizes~\cite{GoremykinaPRL}}. {Though the model is analytically tractable, we can also numerically} access the finite-size behavior of different quantities for {artificially small systems}.  The probability that the system is in the thermal `metallic' phase,  $p_{\rm metal}(L)$,  is shown in  Fig.~\ref{Fig:PhenoRG}. This quantity shows a transition from $p_{\rm metal}(L)\sim 1$ to $p_{\rm metal}(L) \rightarrow 0$ {as a function of a} parameter that tunes transition and that can be interpreted as disorder strength. Interestingly, the step is highly asymmetric: {there is a broad range of} disorder values where {the} system originally appears critical at smaller sizes ($0<p<1$), but then drifts to an ergodic/metallic regime at larger $L$. This asymmetry in this RG model is parametrized by a parameter $\beta$; the limit $\beta\to\infty$  leads to a KT-type RG flow~\cite{GoremykinaPRL,GoremykinaPRL}. However, even for finite values of $\beta$ the numerical data indicates that  relatively large systems of $L\leq 80$ suffer from strong finite-size effects. The apparent collapse of $p_{\rm metal}$ as a function of $W/L$ for $L\leq 18$ breaks down for larger system sizes $L\sim 80$.

\subsection{ Challenges of Finite-size Scaling  in the Localized Phase }
 
For completeness, we briefly remark on finite-size effects on the localized side of the putative MBL transition. Naively, one could expect the finite-size effects to be much weaker at strong disorder, essentially since they are cut off  by the localization length. Indeed, if a localized phase does exist, exact diagonalization studies of small systems can be extremely helpful in extracting its properties, and were instrumental in arriving at the phenomenological description of MBL systems in terms of `localized integrals of motion'. However, we emphasize that this assumes the existence of a localized phase in the first place. In particular, there are situations where numerics can be misleading even in a putative localized regime --- for example, certain systems with long-range interactions --- where  a localized phase can be ruled out on general grounds,\footnote{We note that there is work suggesting that long-range interactions and MBL may be mutually compatible, but these typically do not account for rare-region effects; there are other situations, however, where a direct perturbative calculation indicates that MBL is unstable even at the locator expansion.} but exact diagonalization data looks very similar to that obtained on models thought to host genuine MBL transitions. This indicates that finite-size effects can be subtle even on the localized side. 
 
\section{Does chaos Challenge MBL?\label{sec:vidmar}}

We now turn to a critical examination of Ref.~\onlinecite{Vidmar2019}, which claims absence of MBL at $L\to \infty$ and any disorder $W$ based on two extrapolations of finite-size spectral data. SBPV  studied two models with random on-site fields: the isotropic Heisenberg spin chain (XXX), discussed above, and a XXZ chain with next-nearest neighbor interactions added (denoted as \xxz\ in what follows). The strength of random field required for MBL behavior at finite $L$ differs in the two models: for the former mode, MBL characteristics (such as Poisson level statistics) set in at $W\gtrsim 3.5$, while for the latter model disorder needs to be stronger, $W\gtrsim 8$. This difference stems from the fact that the latter model has longer-range hopping, and kinetic energy is therefore larger, so stronger disorder is needed to localize the systems.

Let us first discuss the better-explored measure, namely the $r$-ratio. They computed the behavior of the $r$-ratio deep in ergodic phase, and as discussed above measured the size of the region $W_{\rm erg}$ (in $W$), {defined as the region where $\langle r\rangle\approx r_{\rm WD}\approx 0.53$}. SBPV found (in agreement with previous studies, and with the models discussed above) that this region $W_{\rm erg} (L)$ grows with $L$, and fitted it with $W_{\rm erg}(L)\propto L$ for $L=12-20$. As a crucial step, SBPV  extrapolated this dependence to  $L\to \infty$, and it was asserted that this is a signature of instability of MBL. As we have demonstrated above, such behaviour of the $r$-ratio is also observed in models where a localized phase is well-established (for instance, compare Figs.~\ref{Fig:RRG-r} and \ref{Fig:Imbrie-r} with Fig.~4 of SBPV). Consequently, it cannot be taken as evidence that an ergodic phase persists to arbitrarily strong disorder in the thermodynamic limit.

Second,  SBPV also considered the relatively less-explored SFF, {see Eq.~(\ref{Eq:Ktau})}. They used this to determine  the Thouless energy, which they then fit to the form
\be\label{eq:Prosen_fit}
E_{\rm Th}(W,L)\sim e^{-W/\Omega} L^{-2},~\text{for}~L\in[12,20],
\ee
 where $\Omega$ is some characteristic energy scale, in the range of $W$ where the system at accessible system sizes is well in the ergodic regime. This scaling ansatz corresponds to assuming  diffusive transport with conductivity scaling as $\sigma(W)\sim e^{-W/\Omega}$.
 
There are several noteworthy points to make about this procedure. First, note that for both models studied ansatz~(\ref{eq:Prosen_fit}) really only works deep in the ergodic phase. For example, for the \xxz\ model, this ansatz works for $1<W<3.5$, whereas MBL behavior is well-developed only at $W>8$ at the sizes accessible by ED. The behavior for the random-field XXX model is similar. Extrapolating this ansatz to the strong-disorder regime is therefore unjustified; SBPV nevertheless assert that it is possible to extrapolate (\ref{eq:Prosen_fit})  to arbitrary disorder and any system  size $L\to \infty$. This {would yield} diffusive transport and an exponentially small but finite conductivity for sufficiently large system sizes $L>L^*(W)$. Now,  it could be argued that perturbative calculations as in Refs.~\cite{Basko06,Mirlin05} might miss  contributions to the  conductivity that depend on disorder as $e^{-W/\Omega}$ 
 as they are non-perturbative in the expansion parameter $1/W$ of the locator expansions on which such calculations are based. However, as we have shown, the class of non-perturbative processes  considered in Ref.~\onlinecite{Imbrie14} {and} further studied by various works on avalanches---and, more generally, any nonperturbative effects based on rare regions---do not give scaling consistent with a critical disorder strength $W_c(L) \propto L$. If the SBPV claims are true, therefore, it seems that they must rely on some hitherto unsuspected nonperturbative instability in \emph{typical} regions. It is difficult to see exactly how to explain these results.   

We therefore conclude that while SBPV provides yet another striking example of the severe finite-size scaling corrections experienced across all extant microscopic numerical investigations of MBL, it does not appear to provide {strong} evidence against the existence of MBL, particularly when viewed in light of the existing numerical studies and theoretical approaches to the transition. We do note that for both models studied the analysis at SBPV is consistent with a transition at higher disorder strength ($W_c\gtrsim 3.5$ for the XXX model, and $W_c\gtrsim 8$ for the \xxz\ model), again consistent with previous scaling analyses~\cite{Alet14, Bera15, PalHuse, Bardarson17}. 
 
\section{Summary and Outlook}
In conclusion, we have demonstrated that several different models, including localization on RRG, phenomenological models, Imbrie-type and random-field XXZ model, exhibit qualitatively similar finite-size effects. In particular, the extent in disorder strength  $W_{\rm erg}(L)$ of the well-ergodic region grows approximately linearly with $L$ at system sizes of up to $L\approx 20$. This behavior does not imply the absence of an MBL phase --- indeed, in all models considered here the existence of localized phase is well-established by analytical means. Extrapolating this scaling to $W, L\to \infty$ (the basic assumption of SBPV) is unjustified and can lead to wrong conclusions. 

The striking similarity of finite-size effects in these ---  {\it a priori} quite different --- models is notable in itself. It would be interesting to find models of MBL which exhibit less severe finite-size effects. A promising direction is to further investigate  experimentally realized models with quasi-periodic potential~\cite{IyerQP}. These were hypothesized to have a different finite-size scaling due to the absence of rare regions, and possibly a distinct universality class of transition~\cite{KhemaniTwoScenarios}. 

We have compared different diagnostics of ergodicity and localization, including the recently-proposed SFF. We note that this quantity
suffers from necessary additional data processing, and from being tailored to the ergodic phase. This measure breaks down with the onset of localization, when $E_{\rm Th}$ becomes of the order of the level spacing. We note that in the single-particle problems there are many other ways to define Thouless energy, which work in both delocalized and localized phases. In the many-body problem, statistics of matrix elements provides one possibility~\cite{Serbyn-16}. In the future, it would be interesting to develop and compare alternative methods for extracting the Thouless energy in many-body systems. Further investigating curvature of levels in response to an external flux (as in~\cite{Thouless72}) may be a promising direction~\cite{Filippone16}. 

Another lesson from our discussion, derived from the results of many previous works, is that it is very difficult to estimate the exact position or critcal properties of the MBL transition based solely on numerical studies.
Developing a complete theory of the MBL transition therefore inevitably requires a combination of rigorous quantum-information bounds, perturbative expansions, and numerics beyond ED. 

%\vspace{1cm}

 \section*{Acknowledgements}
 We thank E.~Altman, D.~Huse, I.M.~Khaymovich, R.~Nandkishore,  S.~Roy, A.~Scardicchio, and S.~Warzel 
 for illuminating discussions. We acknowledge support from EPSRC Grant EP/S020527/1, ``Coherent Many-Body Quantum States of Matter'' (SAP) and US NSF Grant NSF DMR-1653007 (ACP).

\bibliographystyle{apsrev4-1}
\bibliography{mbl}

\end{document}